\documentstyle[12pt]{article}
\begin{document}

\def\ds{\displaystyle}
\def\beq{\begin{equation}}
\def\eeq{\end{equation}}
\def\bea{\begin{eqnarray}}
\def\eea{\end{eqnarray}}
\def\beeq{\begin{eqnarray}}
\def\eeeq{\end{eqnarray}}
\def\ve{\vert}
\def\vel{\left|}
\def\ver{\right|}
\def\nnb{\nonumber}
\def\ga{\left(}
\def\dr{\right)}
\def\aga{\left\{}
\def\adr{\right\}}
\def\lla{\left<}
\def\rra{\right>}
\def\rar{\rightarrow}
\def\nnb{\nonumber}
\def\la{\langle}
\def\ra{\rangle}
\def\ba{\begin{array}}
\def\ea{\end{array}}
\def\tr{\mbox{Tr}}
\def\ssp{{\Sigma^{*+}}}
\def\sso{{\Sigma^{*0}}}
\def\ssm{{\Sigma^{*-}}}
\def\xis0{{\Xi^{*0}}}
\def\xism{{\Xi^{*-}}}
\def\qs{\la \bar s s \ra}
\def\qu{\la \bar u u \ra}
\def\qd{\la \bar d d \ra}
\def\qq{\la \bar q q \ra}
\def\gGgG{\la g^2 G^2 \ra}
\def\q{\gamma_5 \not\!q}
\def\x{\gamma_5 \not\!x}
\def\g5{\gamma_5}
\def\sb{S_Q^{cf}}
\def\sd{S_d^{be}}
\def\su{S_u^{ad}}
\def\ss{S_s^{??}}
\def\sbp{{S}_Q^{'cf}}
\def\sdp{{S}_d^{'be}}
\def\sup{{S}_u^{'ad}}
\def\ssp{{S}_s^{'??}}
\def\sig{\sigma_{\mu \nu} \gamma_5 p^\mu q^\nu}
\def\fo{f_0(\frac{s_0}{M^2})}
\def\ffi{f_1(\frac{s_0}{M^2})}
\def\fii{f_2(\frac{s_0}{M^2})}
\def\O{{\cal O}}
\def\sl{{\Sigma^0 \Lambda}}
\def\es{\!\!\! &=& \!\!\!}
\def\ar{&+& \!\!\!}
\def\ek{&-& \!\!\!}
\def\cp{&\times& \!\!\!}
\def\se{\!\!\! &\simeq& \!\!\!}
\def\kpm{&\pm& \!\!\!}
\def\kmp{&\mp& \!\!\!}


\def\simlt{\stackrel{<}{{}_\sim}}
\def\simgt{\stackrel{>}{{}_\sim}}


\title{
         {\Large
                 {\bf
Explicit expressions of the $\Lambda$ baryon polarizations in
$\Lambda_b \rar \Lambda \ell^+ \ell^-$ decay for the massive 
lepton case 
                 }
         }
      }

\author{\vspace{1cm}\\
{\small T. M. Aliev$^a$ \thanks
{e-mail: taliev@metu.edu.tr}\,\,,
A. \"{O}zpineci$^b$ \thanks
{e-mail: ozpineci@ictp.trieste.it}\,\,,
M. Savc{\i}$^a$ \thanks
{e-mail: savci@metu.edu.tr}} \\
{\small a Physics Department, Middle East Technical University, 
06531 Ankara, Turkey}\\
{\small b  The Abdus Salam International Center for Theoretical Physics,
I-34100, Trieste, Italy} }
\date{}

\begin{titlepage}
\maketitle
\thispagestyle{empty}

\begin{abstract}
We present the explicit form of the expressions of the  
$\Lambda$ baryon polarizations in $\Lambda_b \rar \Lambda \ell^+ \ell^-$ 
decay for the massive lepton case as a complementary to our previous work 
prep: hep--ph/0211447.
\end{abstract}

~~~PACS numbers: 12.60.--i, 13.30.--a
\end{titlepage}

\section*{$\Lambda$ baryon polarizations}

The explicit form of the expressions for the
longitudinal $P_L$ and normal $P_N$ $\Lambda$ baryon polarizations
in $\Lambda_b \rar \Lambda \ell^+ \ell^-$ decay for the massive
lepton case are given as:

\bea
\label{a1}
P_L \es \frac{16 m_{\Lambda_b}^2 \sqrt{\lambda}}
{{\cal T}_0(s) +\frac{1}{3} {\cal T}_2(s)} \Bigg\{
8 m_\ell^2 m_{\Lambda_b}\, \Big(
\mbox{\rm Re}[D_1^\ast E_3 - D_3^\ast E_1] +
\sqrt{r} \mbox{\rm Re}[D_1^\ast D_3 - E_1^\ast E_3)] \Big) \nnb \\
\ek 2 m_\ell m_{\Lambda_b}\, \Big(
(1-\sqrt{r}) \mbox{\rm Re}[(D_1+E_1)^\ast H_2] - 
(1+\sqrt{r}) \mbox{\rm Re}[(D_1-E_1)^\ast F_2]     
\Big) \nnb \\
\ar 16 m_\ell m_{\Lambda_b}\, \Big(
(3 +\sqrt{r}) \mbox{\rm Re}[(A_1-B_1)^\ast C_T f_T] +     
2 (3 -\sqrt{r}) \mbox{\rm Re}[(A_1+B_1)^\ast C_{TE} f_T] \Big)\nnb \\
\ek 16 m_\ell m_{\Lambda_b}^2 (1 -r +2 s) \,
\mbox{\rm Re}[(A_1-B_1)^\ast C_T f_T^V] \nnb \\
\ek 16 m_\ell m_{\Lambda_b}^2 \, \Big\{           
m_{\Lambda_b} (1 -\sqrt{r}) [(1+\sqrt{r})^2 -s]
\mbox{\rm Re}[(A_1-B_1)^\ast C_T f_T^S] \nnb \\
\ek (2 - 2 r +s) \Big( \mbox{\rm Re}[A_2^\ast (C_T - 2 C_{TE}) f_T] -
\mbox{\rm Re}[B_2^\ast (C_T + 2 C_{TE}) f_T] \Big) \Big\} \nnb \\
\ek 2 m_\ell m_{\Lambda_b}^2 s \, \Big\{
\mbox{\rm Re}[(D_3-E_3)^\ast F_2 + (D_3+E_3)^\ast H_2] +
2 m_\ell ( \vel D_3 \ver^2 - \vel E_3 \ver^2 ) \Big\} \nnb \\
\ek 16 m_\ell m_{\Lambda_b}^3 s \, \Big\{
(3-\sqrt{r}) \mbox{\rm Re}[(A_2-B_2)^\ast C_T f_T^V] \nnb \\
\ar m_{\Lambda_b} [(1+\sqrt{r})^2 -s] \mbox{\rm Re}[(A_2-B_2)^\ast C_T f_T^S]
\Big\} \nnb \\
\ek 4 m_{\Lambda_b} (2 m_\ell^2 + m_{\Lambda_b}^2 s) \, 
\mbox{\rm Re}[A_1^\ast B_2 - A_2^\ast B_1] \nnb \\
\ek m_{\Lambda_b}^2 s \, 
\Big( v^2 \mbox{\rm Re}[F_1^\ast H_1] + 
\mbox{\rm Re}[F_2^\ast H_2] \Big) \nnb \\
\ek \frac{4}{3} m_{\Lambda_b}^3 s v^2 \,  
\Big( 3 \mbox{\rm Re}[D_1^\ast E_2 - D_2^\ast E_1] +
\sqrt{r} \mbox{\rm Re}[D_1^\ast D_2 - E_1^\ast E_2] \Big) \nnb \\
\ek \frac{4}{3} m_{\Lambda_b} \sqrt{r} (6 m_\ell^2 + 
m_{\Lambda_b}^2 s v^2) \, \mbox{\rm Re}[A_1^\ast A_2 - B_1^\ast B_2] \nnb \\
\ar \frac{1}{3} \Big\{ 
3 [4 m_\ell^2 + m_{\Lambda_b}^2 (1-r+s)] (\vel A_1 \ver^2 - 
\vel B_1 \ver^2 ) - 3 [4 m_\ell^2 -  m_{\Lambda_b}^2 (1-r+s)] \nnb \\
\cp (\vel D_1 \ver^2 - \vel E_1 \ver^2 ) -  m_{\Lambda_b}^2 (1-r-s) v^2 
(\vel A_1 \ver^2 - \vel B_1 \ver^2 + \vel D_1 \ver^2 - \vel E_1 \ver^2 )
\Big\} \nnb \\
\ek \frac{1}{3} m_{\Lambda_b}^2 \{    
12 m_\ell^2 (1-r) + m_{\Lambda_b}^2 s [3 (1-r+s) + v^2 (1-r-s)] \}
(\vel A_2 \ver^2 - \vel B_2 \ver^2) \nnb \\
\ek \frac{2}{3} m_{\Lambda_b}^4 s (2 - 2 r + s) v^2 \,
(\vel D_2 \ver^2 - \vel E_2 \ver^2) \nnb \\
\ek \frac{128}{3} m_{\Lambda_b}^2 [(1+\sqrt{r})^2 -s]    
\, \Big(
12 m_\ell^2 \mbox{\rm Re}[(C_T f_T^S)^\ast C_{TE} f_T] \nnb \\
\ar 2 m_{\Lambda_b}^2 s v^2 \mbox{\rm Re}[C_T^\ast C_{TE}] 
\mbox{\rm Re}[f_T^\ast f_T^S] \Big)\nnb \\
\ek \frac{128}{3} m_{\Lambda_b}\, \Big\{        
24 m_\ell^2 \mbox{\rm Re}[(C_T f_T^V)^\ast C_{TE} f_T] -
12 m_\ell^2 (1-\sqrt{r}) \mbox{\rm Re}[(C_T f_T)^\ast C_{TE} f_T^V] \nnb \\
\ar 2 m_{\Lambda_b}^2 s [3-\sqrt{r} (3 - 2 v^2)] \mbox{\rm Re}[C_T^\ast C_{TE}] 
\mbox{\rm Re}[f_T^\ast f_T^V] \Big\}\nnb \\
\ar \frac{256}{3} \,    
\{ 6 m_\ell^2 + m_{\Lambda_b}^2 [3 - 3 r - (1-r-s) v^2]\}
\mbox{\rm Re}[C_T^\ast C_{TE}] \vel f_T \ver^2 \Bigg\}~, \nnb 
\eea
\newpage
\bea
\label{a2}
P_N \es \frac{8 \pi m_{\Lambda_b}^3 v \sqrt{s}}
{{\cal T}_0(s) +\frac{1}{3} {\cal T}_2(s)} \Bigg\{
-8 m_\ell m_{\Lambda_b}^2 \lambda \,    
\mbox{\rm Re}[(C_T f_T^S)^\ast (D_1+E_1)] \nnb \\
\ar 8 m_\ell m_{\Lambda_b}^3 \lambda (1+\sqrt{r}) \,
\mbox{\rm Re}[(C_T f_T^S)^\ast (D_2+E_2)] \nnb \\
\ek 2 m_{\Lambda_b} (1-r+s) \sqrt{r} \,
\mbox{\rm Re}[A_1^\ast D_1 + B_1^\ast E_1] \nnb \\
\ar m_\ell [(1-\sqrt{r})^2 -s] \,
\mbox{\rm Re}[(A_1+B_1)^\ast H_1] \nnb \\
\ar m_{\Lambda_b} (1+\sqrt{r}) [(1-\sqrt{r})^2 -s] \, \Big(
4 \mbox{\rm Re}[(C_T f_T)^\ast H_1] - 
8 \mbox{\rm Re}[(C_{TE} f_T)^\ast H_2] \nnb \\
\ek m_\ell \mbox{\rm Re}[(A_2+B_2)^\ast H_1] \Big) \nnb \\
\ar m_{\Lambda_b} (1-\sqrt{r}) [(1+\sqrt{r})^2 -s] \, \Big(
8 \mbox{\rm Re}[(C_{TE} f_T)^\ast F_1] - 
4 \mbox{\rm Re}[(C_T f_T)^\ast F_2] \nnb \\
\ar m_\ell \mbox{\rm Re}[(A_2-B_2)^\ast F_1] \Big) \nnb \\
\ar m_\ell [(1+\sqrt{r})^2 -s] \,
\mbox{\rm Re}[A_1^\ast F_1] \nnb \\
\ar 4 m_{\Lambda_b}^2 s \sqrt{r} \,
\mbox{\rm Re}[A_1^\ast E_2 + A_2^\ast E_1 +B_1^\ast D_2 +
B_2^\ast D_1] \nnb \\
\ek 2 m_{\Lambda_b}^3 s \sqrt{r} (1-r+s) \, 
\mbox{\rm Re}[A_2^\ast D_2 + B_2^\ast E_2^\ast] \nnb \\
\ek 16 m_\ell m_{\Lambda_b} (1+\sqrt{r}) [(1-\sqrt{r})^2 -s] \,
\mbox{\rm Re}[D_1^\ast (C_T+C_{TE}) f_T^V + 
E_1^\ast (C_T-C_{TE}) f_T^V] \nnb \\
\ek 4 m_{\Lambda_b}^2 [(1-\sqrt{r})^2 -s] s \,
\Big( \mbox{\rm Re}[(C_T f_T^V)^\ast H_1] -  
2 \mbox{\rm Re}[(C_{TE} f_T^V)^\ast H_2] \Big) \nnb \\
\ar 16 m_\ell m_{\Lambda_b}^2 [(1-\sqrt{r})^2 -s] s \,
\mbox{\rm Re}[(C_{TE} f_T^V)^\ast (D_3-E_3)] \nnb \\
\ar 2 m_{\Lambda_b} (1-r-s) \, \Big(
\mbox{\rm Re}[A_1^\ast E_1 + B_1^\ast D_1] + 
m_{\Lambda_b}^2 s \mbox{\rm Re}[A_2^\ast E_2 + B_2^\ast D_2] \Big) \nnb \\
\ek m_{\Lambda_b}^2 [(1-r)^2-s^2] \,      
\mbox{\rm Re}[A_1^\ast D_2 + A_2^\ast D_1 + B_1^\ast E_2 + 
B_2^\ast E_1] \nnb \\
\ar 8 m_\ell m_{\Lambda_b}^2 [(1-r)^2-4 \sqrt{r} s - s^2] \,    
\mbox{\rm Re}[(C_T f_T^V)^\ast (D_2 + E_2)] \nnb \\
\ar 8 m_\ell \, \Big\{ 
(1-6 \sqrt{r} + r -s) \mbox{\rm Re}[(C_T f_T)^\ast E_1] +
2 (1+6 \sqrt{r} + r -s) \mbox{\rm Re}[(C_{TE} f_T)^\ast E_1] \nnb \\
\ek 4 m_{\Lambda_b} (1 - \sqrt{r}) [(1+\sqrt{r})^2 -s]
\mbox{\rm Re}[(C_{TE} f_T)^\ast D_2] \nnb \\
\ek 2 m_{\Lambda_b} (1 + \sqrt{r}) [(1-\sqrt{r})^2 -s]
\mbox{\rm Re}[(C_T f_T)^\ast D_2] \Big\} \nnb \\
\ek m_\ell [(1+\sqrt{r})^2 -s] \,
\mbox{\rm Re}[B_1^\ast F_1] \nnb \\
\ek 8 m_\ell m_{\Lambda_b} \, \Big\{ (1 - \sqrt{r}) 
[(1 + \sqrt{r})^2 - s] \mbox{\rm Re}[(C_T f_T)^\ast E_3] \nnb \\
\ek 2 (1 + \sqrt{r}) [(1 - \sqrt{r})^2 - s] 
\mbox{\rm Re}[(C_{TE} f_T)^\ast E_3] \Big\} \nnb \\
\ar 16 m_\ell m_{\Lambda_b} \, \Big\{
2 (1 - \sqrt{r}) [(1 + \sqrt{r})^2 - s] 
\mbox{\rm Re}[(C_{TE} f_T)^\ast E_2] \nnb \\
\ek (1 + \sqrt{r}) [(1 - \sqrt{r})^2 - s] \,
\mbox{\rm Re}[(C_T f_T)^\ast E_2] \Big\} \nnb \\
\ek 8 m_\ell \, \Big\{
2 (1 + 6 \sqrt{r} + r - s) \mbox{\rm Re}[(C_{TE} f_T)^\ast D_1] -
(1 - 6 \sqrt{r} + r - s) \mbox{\rm Re}[(C_T f_T)^\ast D_1]
\Big\} \nnb \\
\ek 8 m_\ell m_{\Lambda_b} \, \Big\{
2 (1 + \sqrt{r}) [(1 - \sqrt{r})^2 - s]
\mbox{\rm Re}[(C_{TE} f_T)^\ast D_3] \nnb \\
\ar (1 - \sqrt{r}) [(1 + \sqrt{r})^2 - s]
\mbox{\rm Re}[(C_T f_T)^\ast D_3]\Bigg\}~. \nnb
\eea

\end{document}